\documentstyle[12pt]{article}
\def\D{\Delta}
\def\Ph{\phi}

\def\ka{\kappa}

\def\vphi{\varphi}

\def\z{\zeta}
\def\spa{\sigma_{\parallel}}
\def\spe{\sigma_{\perp}}
\title{Functional Integration in Bose Systems with Hard-Core Interaction}
\author{\large Klaus Ziegler \\[3mm]
\em Max-Planck-Institut f\"ur Physik Komplexer Systeme,  \\
\em Au\ss enstelle Stuttgart, \\
\em D-70506 Stuttgart, Germany}
\date{}
\begin{document}
\maketitle
\begin{abstract}
%Here there will be an Abstract.
A grand canonical ensemble of interacting bosons is considered. The zero
temperature phase diagram is evaluated from the mean-field approximation
of the functional integral. Three phases are found: a superfluid, a normal
fluid and a Mott insulator.
\end{abstract}
Quantum systems with many-body interactions are an important application of
functional integrals. As an example a model of interacting bosons is
considered. It is defined by the
Hamiltonian
\begin{equation}
H=-\sum_{<r,r'>}J_{r,r'}\Phi_r^\dagger\Phi_{r'}-\mu\sum_{r}\Phi_r^\dagger\Phi_r
+{V\over 2}\sum_{r}\Phi_r^\dagger\Phi_r\Phi_{r}^\dagger\Phi_{r},
\end{equation}
where $\Phi_r^\dagger$ ($\Phi_r$) creates (annihilates) a boson at the lattice
site $r$. The first term describes bosons hopping from site $r'$ to a nearest
neighbor site $r$. $\mu$ is a single particle potential and $V$ the coupling
constant of an on-site interaction.

Following the standard procedure (e.g. Ref.\cite{negele}) the corresponding
functional integral for a grand canonical ensemble of bosons reads
\begin{equation}
Z=\int\exp(-S){\cal D}[\phi,\phi^*]
\label{z}
\end{equation}
with the action
\begin{equation}
S=\D\sum_{r,t}{1\over \D}\phi^*(r,t)[\phi(r,t)-\phi(r,t-\D)]
+\D\sum_t {1\over\hbar}H[\phi^*(r,t),\phi(r,t-\D)].
\end{equation}
We have introduced a discrete imaginary time with interval length $\D$ and a
complex field $\Ph$ with periodic boundaries in time
%\begin{equation}
$\Ph(t=0,r)=\Ph(t=\beta\hbar,r)$.
%\end{equation}
The parameter $\mu$ plays in (\ref{z}) the role of the chemical potential.

%The lattice points $r$ will be denoted as {\it sites}
%in contrast to the points $(r,t)$ of the space-time lattice which will be
%called {\it points}.

The local part of the Hamiltonian in $S$ can be approximated
$\phi^*(r,t)\phi(r,t-\D)\approx\phi(r,t)\phi^*(r,t)=|\phi(r,t)|^2$, neglecting
terms of $O(\D^2)$ in $S$.\cite{negele}
The interaction term has a positive coupling constant $V$ (repulsion).
Thermal fluctuations are switched off by taking implicitly the limit 
$\beta\to\infty$.
A random path expansion \cite{glimm} can be created by extracting the local
part from the action

\begin{equation}
S_0=-\D\sum_x\Big[(\mu-1/\D)n_x-{V\over2}n_x^2\Big],
\end{equation}
where $n_x=|\Ph_x|^2$ and $\hbar=1$. The expansion in terms of 
$S_1\equiv S-S_0$
generates locally powers of the complex field $\Ph_x$ and $\Ph_x^*$. 
Since $S_0$ does not depend on the phase of $\Ph_x$ we obtain from the
integration w.r.t. $\Ph$ nonzero
quantities only if $\Ph_x$ and $\Ph_x^*$ appear with the same power at each
point $x=(t,r)$
\begin{equation}
\int\Ph_x^{l_x}{\Ph_x^*}^{l_x^*}\exp\Big\{\D\big[(\mu-1/\D)n_x-
{V\over2}n^2_x\big]\Big\} d\Ph_x^* d\Ph_x=g_{l_x}\delta_{l_x,l_{x}^*}.
\end{equation}
Thus, the random path expansion of the partition function leads to a
grand-canonical ensemble of Bose world lines (BWLs), going from $t=0$ to
$t=\beta$, which are directed along the $t$-axis. The interaction
of the bosons corresponds to the coefficient $g_l$ at $x$ which counts
$l$ visiting bosons at $r$ at the time $t$. The statistics
of the BWLs simplifies essentially if we apply a
hard-core approximation by introducing the restriction that a point $x$
can be visited at most by one boson
\begin{equation}
g_l=0 {\ \ \ \ \ \ \rm for\ \ } l\ge 2.
\end{equation}
This condition projects onto boson realizations with hard-core interaction
at each point. 
%The hard-core implies also that the random path expansion of $Z$ 
%terminates on a finite lattice.
The contributions to the expansion are empty points with weight
$g_0\equiv\zeta$ and world line elements which connect points
$x$ and $x'$ with statistical weight
\begin{equation}
w_{x,x'}\equiv w_{x-x'}=\cases{
\D g_1J_{r,r'}\equiv\alpha & for $x'=x+e'_\mu$\cr
1                          & for $x'=x+\D e_t$\cr
0                          & otherwise\cr
}.
\end{equation}
$e_\mu$ are the unit vectors in the $d$-dimensional space, $e_t$ is the unit
vector in time and $e'_\mu=e_\mu+\D e_t$. This model has only two independent
parameters, $\alpha$ and $\zeta$.
%%%%%%%%%%%%%%%%%%%%%%%%%%%%%%%%%%%%%%%%%%%%%%%%%%%%%%%%%%%%%%%%%%%%%%%%%% 
%The continuous time limit $\D\to0$ can be understood in terms of Brownian
%motion: the lattice constant must scale like $\sqrt{\D}$ (and therefore,
%$J_{r,r'}$ like $\D^{-1}$) in order to get a diffusion process for 
%non-interacting bosons.
%%%%%%%%%%%%%%%%%%%%%%%%%%%%%%%%%%%%%%%%%%%%%%%%%%%%%%%%%%%%%%%%%%%%%%%%%%%
We may write for the partition function
of a system on a lattice with ${\cal N}$ sites
\begin{equation}
Z_{h.-c.}
=\sum_{n=0}^{\cal N}{\zeta^{({\cal N}-n)\beta}\over n!}\sum_{\{r_j(t)\}_n\in
I_n} P(\{ r_j(t)\}_n),
\label{zhc}
\end{equation}
where $P(\{ r_j(t)\}_n)$ is the weight of a given configuration
$\{ r_j(t)\}_n$ of $n$ BWLs
\begin{equation}
P(\{ r_j(t)\}_n)=\prod_{x,x'\in\{r_j(t)\}_n}w_{x,x'}.
\end{equation}
%where $\{ r_j(t)\}_n$ has to respect the hard-core condition.
$I_n$ are all configurations of BWLs going from $t=0$ to $t=\beta$
which are allowed by the hard-core condition. It has been shown that the
partition function of hard-core
bosons (\ref{zhc}) can be rewritten for dimensions $d>1$ in terms of
interacting fermions
\cite{zie1,zie2,zie3}
\begin{equation}
S_{\rm ferm}=\sum_{x,x'}(w_{x,x'}+\zeta\delta_{x,x'})
\psi^1_x\psi^2_x{\bar\psi}^1_{x'}{\bar\psi}^2_{x'},
\end{equation}
with Grassmann fields $\psi^a_x$.
The hard-core interaction is now represented by the exclusion (Pauli)
principle of the fermions. Especially in $d=1$, the hard-core bosons
cannot exchange their positions. This implies that an additional interaction
between the fermions is not necessary, and the hard-core bosons are equivalent
to {\it non-interacting} fermions \cite{lieb}.

The hard-core boson model has been generalized to a model of bosons with $N$
colors \cite{zie2} or a model with $N$ levels at each point \cite{zie3}.
In the latter the bosons can occupy the $N$ levels statistically. In both
cases the limit $N\to\infty$ can be solved exactly. The formal reason is a
formation of composite fields from pairs of fermions.
%which couple linearly to non-interacting fermions.
For the $N$ color bosons these are nonlocal fields and in the case of the $N$
bosonic levels these are local fields. The composite fields do not depend on
the number $N$ but $N$ appears as a prefactor of the action. Therefore, a
saddle point integration can be performed for $N\to\infty$ in both cases.
The results of the calculations of Refs.\cite{zie2,zie3} will be used in the
following to derive the phase diagram of hard-core bosons at zero temperature.
\\

\noindent
a) {\it local order parameter}: Superfluid (SF)

The effective functional integral of the $N$ level model of Ref.\cite{zie3}
is related to a local complex order parameter field $(\vphi_x,\chi_x)$. 
The corresponding partition function reads
\begin{equation}
Z_{\rm loc}=\int\exp\Big\{ -N\big[(\vphi,v_1^{-1}\vphi^*)+(\chi,\chi^*)-
\sum_x\log [\z +(\vphi_x+i\chi_x)(\vphi_x^*+i\chi_x^*)]\big]\Big\}
%\prodd\vphi_x d\chi_x
{\cal D}[\vphi,\chi],
\label{zl}
\end{equation}
where $v_1$ is the matrix $w+1$.
There are two mean-field solutions for the saddle point equation
$\delta s_{\rm loc}=0$: a trivial one with $\vphi=\chi=0$ and a nontrivial 
one with $\vphi=-i\chi=2\sqrt{1-\zeta}$. 
The stability of the two solutions is found from the fluctuations around the
saddle point \cite{zie3}: the trivial solution is stable for $\zeta>\alpha^4$
and the nontrivial one for $0\le\zeta\le\alpha^4$.
Thus the critical point is $\zeta_c=\alpha^4$.
The mean-field approximation of the action is $-\log\zeta$ for the trivial
solution and $\alpha^2-4\log\alpha-\zeta/\alpha^2$ for the nontrivial
solution.
The fluctuations are gapless for $0\le\zeta\le\alpha^4$ due to Goldstone
bosons created by a spontaneously broken $U(1)$ symmetry. The corresponding
phase is a superfluid. This result can also be obtained if a
Ginzburg-Landau expansion is applied directly to the action in (\ref{zl}). The
expansion should be valid for small density of bosons. This is very
similar to the discussion of weakly interacting bosons \cite{popov}.
However, this mean-field solution is not a complete description of the physical states
of hard-core bosons on a lattice if higher densities are included. This can
be seen if a nonlocal order parameter is considered.\\

\noindent
b) {\it nonlocal order parameter}: Normal Fluid (NF) and Mott Insulator (MI)

It is possible to relate an order parameter field to the elements of the BWLs
by replacing $w_{x,x'}\to u_{x,x'}$
% by $u_{x,x+\D e_t}$ and $u_{x,x+e'_\mu}$
\cite{zie1,zie2}. The partition function then reads
\begin{equation}
Z_{\rm nonloc}=\int\exp\Big\{ -N\sum_{x,x'}{(u_{x,x'})^2\over w_{x,x'}}
%{\rm Tr}(uu^T)
+2N\log\det (\sqrt{\z} + u)\Big\}
%\prod du_{x,x'}
{\cal D}[u].
\end{equation}
Two different contributions can be distinguished,
one is the longitudinal component $u_{x,x+\D e_t}$ and the other
the transverse component $u_{x,x+e'_\mu}$. A mean-field approximation
$u_{x,x+\D e_t}\approx\spa$ and $u_{x,x+e'_\mu}\approx\spe$ has to satisfy the
saddle point conditions
\begin{equation}
{\partial s_0\over\partial\spa}=\spa-{2\over\spa}I=0,
{\ \ \ }
{\partial s_0\over\partial\spe}=2d\spe-{2\over\spe}(n-I)=0
\label{spe}
\end{equation}
with the integrals
%\begin{equation}
$I=\spa\int_{-1}^1\Theta ({\ka'}^2-\zeta)/ka'\rho (\ka)d\ka$ and
%{\ \ \ }
$n=\int_{-1}^1\Theta ({\ka'}^2-\zeta)\rho (\ka)d\ka$.
%\end{equation}
$n$ is the density of BWLs, $\kappa'=2d\alpha\spe\kappa+\spa$ and
$\rho (\kappa)=\int\delta[(1/d)\sum_{j=1}^{d}
\cos k_j - \kappa ] dk_1\ldots dk_d/(2\pi)^d$.
$\Theta$ is the Heaviside step function. The mean-field action reads
\begin{equation}
s_0= d\spe^2+{1\over 2}\spa^2-\log\zeta
-\int_{-1}^1\log({\ka'}^2/\zeta)\Theta ({\ka'}^2-\zeta)\rho(\ka)d\ka-i\pi.
\end{equation}
The saddle point conditions (\ref{spe}) imply $\spa^2+2d\spe^2=2n$.
There are three different solutions: $(\spa,\spe)=(0,0)$ (empty phase),
$(0,\sqrt{n/d})$ (NF) and $(\sqrt{2n},0)$ (MI).
The approximation $\rho(\kappa)\approx(1/2)\Theta(1-|\kappa|)$
gives for the mean-field action the values shown in Table 1,
%\begin{equation}
%s_{\rm nonloc}=\cases{
%1-\log2                       & for $\spa=\sqrt{2n}$, $\spe=0$ \cr
%3n-\log n-2\log (2\sqrt{d}\alpha) & for $\spa=0$, $\spe=\sqrt{n/d}$\cr
%-\log\zeta                  & for $\spa=0$, $\spe=0$\cr
%},
%\end{equation}
where $\zeta=4d(1-n)^2n\alpha^2$ for the NF. Although the SF of
a) is also a phase with fluctuating BWLs, the NF can be distinguished by
the fact that its fluctuations have a gap in contrast to the gapless
fluctuations of the SF. Moreover, the SF has a nonzero winding number
in contrast to the zero winding number of the NF. The fluctuations of the
MI also have a gap which is related to an incompressible array of BWLs. The
latter is commensurate with the lattice. An additional nonlocal interaction
leads to many commensurate MI phases \cite{fei}.\\

\noindent
{\it Discussion}: For large values of $\zeta$ and $T=0$ there will be always
an empty phase (i.e., no bosons). (In the case of nonzero temperatures this
phase would be a NF due to thermal fluctuations.)
The parameter $\alpha$ is the hopping rate of the bosons.
Starting from the empty phase at zero temperature the density can be
increased by a reduction of $\zeta$.
Depending on the values of $\alpha$ the NF (small $\alpha$) or the
SF (large $\alpha$) will be reached. An increasing density will eventually
destroy the SF due to strong collisions between the bosons,
and a transition to a NF will take place. Finally, even the NF will be
destroyed: a state can be formed which is commensurate with the lattice,
provided $\alpha$ is not too large. This is the MI.
%There is no MI if $\alpha>e/\sqrt{2d}$. Moreover, there is no MI for 
%$\zeta>2/e\approx0.86$. There is no NF if $\zeta>16\alpha^2d/27$.
%Finally, there is no SF if $\zeta>\alpha^4$.
As an examples the phase diagram is shown for $d=3$ in Fig.1.\\

Table 1: The mean-field solutions for local and nonlocal order parameters

\begin{tabular}{c c c c}
\hline
 phase:          &  Empty Phase  & Fluid Phase & Mott Insulator \\
\hline
 density:        &    $n=0$      &    $0<n<1$   &      $n=1$       \\
 $\spa$:          &       0       &     0        &   $\sqrt{2n}$   \\
 $\spe$:          &       0       &$\sqrt{n/d}$  &    0            \\
$s_{\rm nonloc}$: & $-\log\zeta$ & $3n-\log n-2\log(2\sqrt{d}\alpha)$ (NF)
&$1-\log2$\\
 $\vphi$:         &       0       &  $2\sqrt{1-\zeta/\alpha^4}$ &   --  \\ 
$s_{\rm loc}$: & $-\log\zeta$  & $\alpha^2-\zeta/\alpha^2-4\log\alpha$ (SF)
&  -- \\
\end{tabular}
\vfill
\pagebreak

\begin{figure}
\begin{center}
% GNUPLOT: LaTeX picture

\setlength{\unitlength}{0.240900pt}

\ifx\plotpoint\undefined\newsavebox{\plotpoint}\fi

\sbox{\plotpoint}{\rule[-0.200pt]{0.400pt}{0.400pt}}%

\begin{picture}(1500,900)(0,0)

\font\gnuplot=cmr10 at 10pt

\gnuplot

\sbox{\plotpoint}{\rule[-0.200pt]{0.400pt}{0.400pt}}%

\put(220.0,113.0){\rule[-0.200pt]{0.400pt}{184.048pt}}

\put(220.0,113.0){\rule[-0.200pt]{4.818pt}{0.400pt}}

\put(198,113){\makebox(0,0)[r]{0.7}}

\put(1416.0,113.0){\rule[-0.200pt]{4.818pt}{0.400pt}}

\put(220.0,208.0){\rule[-0.200pt]{4.818pt}{0.400pt}}

\put(198,208){\makebox(0,0)[r]{0.8}}

\put(1416.0,208.0){\rule[-0.200pt]{4.818pt}{0.400pt}}

\put(220.0,304.0){\rule[-0.200pt]{4.818pt}{0.400pt}}

\put(198,304){\makebox(0,0)[r]{0.9}}

\put(1416.0,304.0){\rule[-0.200pt]{4.818pt}{0.400pt}}

\put(220.0,399.0){\rule[-0.200pt]{4.818pt}{0.400pt}}

\put(198,399){\makebox(0,0)[r]{1}}

\put(1416.0,399.0){\rule[-0.200pt]{4.818pt}{0.400pt}}

\put(220.0,495.0){\rule[-0.200pt]{4.818pt}{0.400pt}}

\put(198,495){\makebox(0,0)[r]{1.1}}

\put(1416.0,495.0){\rule[-0.200pt]{4.818pt}{0.400pt}}

\put(220.0,591.0){\rule[-0.200pt]{4.818pt}{0.400pt}}

\put(198,591){\makebox(0,0)[r]{1.2}}

\put(1416.0,591.0){\rule[-0.200pt]{4.818pt}{0.400pt}}

\put(220.0,686.0){\rule[-0.200pt]{4.818pt}{0.400pt}}

\put(198,686){\makebox(0,0)[r]{1.3}}

\put(1416.0,686.0){\rule[-0.200pt]{4.818pt}{0.400pt}}

\put(220.0,782.0){\rule[-0.200pt]{4.818pt}{0.400pt}}

\put(198,782){\makebox(0,0)[r]{1.4}}

\put(1416.0,782.0){\rule[-0.200pt]{4.818pt}{0.400pt}}

\put(220.0,877.0){\rule[-0.200pt]{4.818pt}{0.400pt}}

\put(198,877){\makebox(0,0)[r]{1.5}}

\put(1416.0,877.0){\rule[-0.200pt]{4.818pt}{0.400pt}}

\put(220.0,113.0){\rule[-0.200pt]{0.400pt}{4.818pt}}

\put(220,68){\makebox(0,0){0}}

\put(220.0,857.0){\rule[-0.200pt]{0.400pt}{4.818pt}}

\put(394.0,113.0){\rule[-0.200pt]{0.400pt}{4.818pt}}

\put(394,68){\makebox(0,0){0.2}}

\put(394.0,857.0){\rule[-0.200pt]{0.400pt}{4.818pt}}

\put(567.0,113.0){\rule[-0.200pt]{0.400pt}{4.818pt}}

\put(567,68){\makebox(0,0){0.4}}

\put(567.0,857.0){\rule[-0.200pt]{0.400pt}{4.818pt}}

\put(741.0,113.0){\rule[-0.200pt]{0.400pt}{4.818pt}}

\put(741,68){\makebox(0,0){0.6}}

\put(741.0,857.0){\rule[-0.200pt]{0.400pt}{4.818pt}}

\put(915.0,113.0){\rule[-0.200pt]{0.400pt}{4.818pt}}

\put(915,68){\makebox(0,0){0.8}}

\put(915.0,857.0){\rule[-0.200pt]{0.400pt}{4.818pt}}

\put(1089.0,113.0){\rule[-0.200pt]{0.400pt}{4.818pt}}

\put(1089,68){\makebox(0,0){1}}

\put(1089.0,857.0){\rule[-0.200pt]{0.400pt}{4.818pt}}

\put(1262.0,113.0){\rule[-0.200pt]{0.400pt}{4.818pt}}

\put(1262,68){\makebox(0,0){1.2}}

\put(1262.0,857.0){\rule[-0.200pt]{0.400pt}{4.818pt}}

\put(1436.0,113.0){\rule[-0.200pt]{0.400pt}{4.818pt}}

\put(1436,68){\makebox(0,0){1.4}}

\put(1436.0,857.0){\rule[-0.200pt]{0.400pt}{4.818pt}}

\put(220.0,113.0){\rule[-0.200pt]{292.934pt}{0.400pt}}

\put(1436.0,113.0){\rule[-0.200pt]{0.400pt}{184.048pt}}

\put(220.0,877.0){\rule[-0.200pt]{292.934pt}{0.400pt}}

\put(45,495){\makebox(0,0){$\alpha$}}

\put(828,23){\makebox(0,0){$\sqrt{\zeta}$}}

\put(567,638){\makebox(0,0)[l]{NORMAL FLUID}}

\put(481,256){\makebox(0,0)[l]{MOTT INSULATOR}}

\put(1219,590){\makebox(0,0)[l]{SUPERFLUID}}

\put(1175,304){\makebox(0,0)[l]{EMPTY}}

\put(220.0,113.0){\rule[-0.200pt]{0.400pt}{184.048pt}}

\sbox{\plotpoint}{\rule[-0.400pt]{0.800pt}{0.800pt}}%

\put(246,495){\usebox{\plotpoint}}

\multiput(246.00,493.09)(1.459,-0.501){185}{\rule{2.525pt}{0.121pt}}

\multiput(246.00,493.34)(273.759,-96.000){2}{\rule{1.263pt}{0.800pt}}

\multiput(525.00,397.09)(1.528,-0.501){183}{\rule{2.634pt}{0.121pt}}

\multiput(525.00,397.34)(283.534,-95.000){2}{\rule{1.317pt}{0.800pt}}

\multiput(814.00,302.09)(1.570,-0.501){183}{\rule{2.701pt}{0.121pt}}

\multiput(814.00,302.34)(291.394,-95.000){2}{\rule{1.351pt}{0.800pt}}

\multiput(1111.00,207.09)(1.556,-0.502){89}{\rule{2.667pt}{0.121pt}}

\multiput(1111.00,207.34)(142.465,-48.000){2}{\rule{1.333pt}{0.800pt}}

\put(1046,237){\usebox{\plotpoint}}

\multiput(1047.40,237.00)(0.526,0.738){7}{\rule{0.127pt}{1.343pt}}

\multiput(1044.34,237.00)(7.000,7.213){2}{\rule{0.800pt}{0.671pt}}

\multiput(1054.40,247.00)(0.520,0.554){9}{\rule{0.125pt}{1.100pt}}

\multiput(1051.34,247.00)(8.000,6.717){2}{\rule{0.800pt}{0.550pt}}

\multiput(1062.41,256.00)(0.505,0.661){37}{\rule{0.122pt}{1.255pt}}

\multiput(1059.34,256.00)(22.000,26.396){2}{\rule{0.800pt}{0.627pt}}

\multiput(1084.41,285.00)(0.507,0.593){25}{\rule{0.122pt}{1.150pt}}

\multiput(1081.34,285.00)(16.000,16.613){2}{\rule{0.800pt}{0.575pt}}

\multiput(1100.41,304.00)(0.501,0.710){127}{\rule{0.121pt}{1.334pt}}

\multiput(1097.34,304.00)(67.000,92.231){2}{\rule{0.800pt}{0.667pt}}

\put(1166,399){\usebox{\plotpoint}}

\multiput(1167.41,399.00)(0.511,4.288){17}{\rule{0.123pt}{6.600pt}}

\multiput(1164.34,399.00)(12.000,82.301){2}{\rule{0.800pt}{3.300pt}}

\multiput(1179.38,495.00)(0.560,2.775){3}{\rule{0.135pt}{3.240pt}}

\multiput(1176.34,495.00)(5.000,12.275){2}{\rule{0.800pt}{1.620pt}}

\multiput(1184.40,514.00)(0.516,1.747){11}{\rule{0.124pt}{2.778pt}}

\multiput(1181.34,514.00)(9.000,23.235){2}{\rule{0.800pt}{1.389pt}}

\multiput(1193.40,543.00)(0.526,1.526){7}{\rule{0.127pt}{2.371pt}}

\multiput(1190.34,543.00)(7.000,14.078){2}{\rule{0.800pt}{1.186pt}}

\multiput(1200.40,562.00)(0.512,1.387){15}{\rule{0.123pt}{2.309pt}}

\multiput(1197.34,562.00)(11.000,24.207){2}{\rule{0.800pt}{1.155pt}}

\multiput(1211.40,591.00)(0.516,1.116){11}{\rule{0.124pt}{1.889pt}}

\multiput(1208.34,591.00)(9.000,15.080){2}{\rule{0.800pt}{0.944pt}}

\multiput(1220.41,610.00)(0.508,0.953){23}{\rule{0.122pt}{1.693pt}}

\multiput(1217.34,610.00)(15.000,24.485){2}{\rule{0.800pt}{0.847pt}}

\multiput(1235.40,638.00)(0.514,0.988){13}{\rule{0.124pt}{1.720pt}}

\multiput(1232.34,638.00)(10.000,15.430){2}{\rule{0.800pt}{0.860pt}}

\multiput(1245.41,657.00)(0.507,0.866){27}{\rule{0.122pt}{1.565pt}}

\multiput(1242.34,657.00)(17.000,25.752){2}{\rule{0.800pt}{0.782pt}}

\multiput(1262.41,686.00)(0.501,0.744){121}{\rule{0.121pt}{1.388pt}}

\multiput(1259.34,686.00)(64.000,92.120){2}{\rule{0.800pt}{0.694pt}}

\multiput(1326.41,781.00)(0.503,0.649){67}{\rule{0.121pt}{1.238pt}}

\multiput(1323.34,781.00)(37.000,45.431){2}{\rule{0.800pt}{0.619pt}}

\multiput(1363.40,829.00)(0.520,0.627){9}{\rule{0.125pt}{1.200pt}}

\multiput(1360.34,829.00)(8.000,7.509){2}{\rule{0.800pt}{0.600pt}}

\multiput(1371.40,839.00)(0.520,0.554){9}{\rule{0.125pt}{1.100pt}}

\multiput(1368.34,839.00)(8.000,6.717){2}{\rule{0.800pt}{0.550pt}}

\put(1172,444){\usebox{\plotpoint}}

\multiput(1172.00,445.39)(1.020,0.536){5}{\rule{1.667pt}{0.129pt}}

\multiput(1172.00,442.34)(7.541,6.000){2}{\rule{0.833pt}{0.800pt}}

\multiput(1183.00,451.39)(1.132,0.536){5}{\rule{1.800pt}{0.129pt}}

\multiput(1183.00,448.34)(8.264,6.000){2}{\rule{0.900pt}{0.800pt}}

\multiput(1195.00,457.39)(1.020,0.536){5}{\rule{1.667pt}{0.129pt}}

\multiput(1195.00,454.34)(7.541,6.000){2}{\rule{0.833pt}{0.800pt}}

\multiput(1206.00,463.39)(1.132,0.536){5}{\rule{1.800pt}{0.129pt}}

\multiput(1206.00,460.34)(8.264,6.000){2}{\rule{0.900pt}{0.800pt}}

\multiput(1218.00,469.39)(1.020,0.536){5}{\rule{1.667pt}{0.129pt}}

\multiput(1218.00,466.34)(7.541,6.000){2}{\rule{0.833pt}{0.800pt}}

\multiput(1229.00,475.39)(1.020,0.536){5}{\rule{1.667pt}{0.129pt}}

\multiput(1229.00,472.34)(7.541,6.000){2}{\rule{0.833pt}{0.800pt}}

\multiput(1240.00,481.38)(1.600,0.560){3}{\rule{2.120pt}{0.135pt}}

\multiput(1240.00,478.34)(7.600,5.000){2}{\rule{1.060pt}{0.800pt}}

\multiput(1252.00,486.39)(1.020,0.536){5}{\rule{1.667pt}{0.129pt}}

\multiput(1252.00,483.34)(7.541,6.000){2}{\rule{0.833pt}{0.800pt}}

\multiput(1263.00,492.39)(1.132,0.536){5}{\rule{1.800pt}{0.129pt}}

\multiput(1263.00,489.34)(8.264,6.000){2}{\rule{0.900pt}{0.800pt}}

\multiput(1275.00,498.39)(1.020,0.536){5}{\rule{1.667pt}{0.129pt}}

\multiput(1275.00,495.34)(7.541,6.000){2}{\rule{0.833pt}{0.800pt}}

\multiput(1286.00,504.38)(1.600,0.560){3}{\rule{2.120pt}{0.135pt}}

\multiput(1286.00,501.34)(7.600,5.000){2}{\rule{1.060pt}{0.800pt}}

\multiput(1298.00,509.39)(1.020,0.536){5}{\rule{1.667pt}{0.129pt}}

\multiput(1298.00,506.34)(7.541,6.000){2}{\rule{0.833pt}{0.800pt}}

\multiput(1309.00,515.38)(1.432,0.560){3}{\rule{1.960pt}{0.135pt}}

\multiput(1309.00,512.34)(6.932,5.000){2}{\rule{0.980pt}{0.800pt}}

\multiput(1320.00,520.39)(1.132,0.536){5}{\rule{1.800pt}{0.129pt}}

\multiput(1320.00,517.34)(8.264,6.000){2}{\rule{0.900pt}{0.800pt}}

\multiput(1332.00,526.39)(1.020,0.536){5}{\rule{1.667pt}{0.129pt}}

\multiput(1332.00,523.34)(7.541,6.000){2}{\rule{0.833pt}{0.800pt}}

\multiput(1343.00,532.38)(1.600,0.560){3}{\rule{2.120pt}{0.135pt}}

\multiput(1343.00,529.34)(7.600,5.000){2}{\rule{1.060pt}{0.800pt}}

\multiput(1355.00,537.39)(1.020,0.536){5}{\rule{1.667pt}{0.129pt}}

\multiput(1355.00,534.34)(7.541,6.000){2}{\rule{0.833pt}{0.800pt}}

\multiput(1366.00,543.38)(1.600,0.560){3}{\rule{2.120pt}{0.135pt}}

\multiput(1366.00,540.34)(7.600,5.000){2}{\rule{1.060pt}{0.800pt}}

\end{picture}

\end{center}
\end{figure}
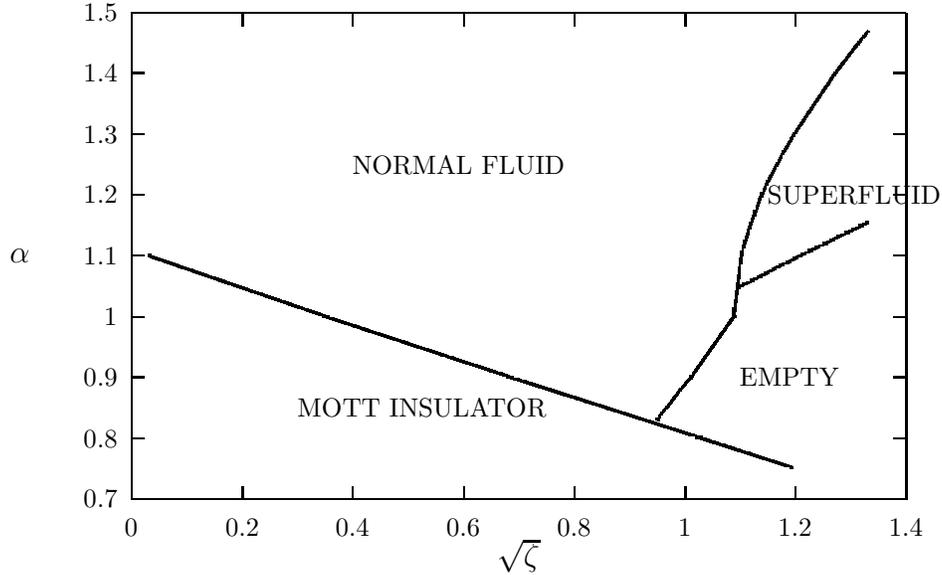
Fig1: $T=0$ phase diagram for hard-core bosons. $\alpha$ is the
hopping rate of the bosons and $\sqrt\zeta$ the fugacity of empty
sites. Thermal fluctuations ($T>0$) would destroy the empty phase and
change it into a NF.

\footnotesize{

}
\end{document}